\documentclass[11pt,conference]{IEEEtran}
\usepackage{graphics}
\usepackage{amssymb}
\usepackage{amscd}
\usepackage{amsmath}
\usepackage{epsfig}
\usepackage{rotating}
\usepackage{amsmath}
\usepackage{amsfonts}
\usepackage{url}
\usepackage[english]{babel}
\renewcommand{\baselinestretch}{0.89}
\begin{document}
\title{A Game Theoretic Framework for Decentralized Power Allocation in IDMA Systems}
\author{\IEEEauthorblockN{Samir Medina Perlaza}
\IEEEauthorblockA{France Telecom R\&D - Orange Labs, France\\
samir.medinaperlaza@orange-ftgroup.com}\\
\and
\IEEEauthorblockN{Laura Cottatellucci}
\IEEEauthorblockA{ Institute Eurecom, France\\
Laura.Cottatellucci@eurecom.fr}
\and
\IEEEauthorblockN{M\'erouane Debbah}
\IEEEauthorblockA{SUPELEC, France\\
Merouane.Debbah@supelec.fr} }
\maketitle
\begin{abstract}
\boldmath
In this contribution we present a decentralized power allocation algorithm for the uplink interleave division multiple access (IDMA) channel. Within the proposed optimal strategy for power allocation, each user aims at  selfishly maximizing its own utility function. An iterative chip-by-chip (CBC) decoder at the receiver and a rational selfish behavior of all the users according to a classical game-theoretical framework are the underlying assumptions of this work. This approach leads to a channel inversion policy where the optimal power level is set locally at each terminal based on the knowledge of its own channel realization, the noise level at the receiver and the number of active users in the network.
\end{abstract}
\section{Introduction}
Interleave division multiple access (IDMA) has been identified as
a promising multiple access technique in the context of cellular
networks \cite{Ping2006TransWireless} and self-organizing networks
(e.g \emph{ad hoc networks}) \cite{Kusume2006VTC}. In this domain,
distributed power algorithms play a central practical and
theoretical role. Nonetheless, the study of \emph{distributed}
power allocation algorithms for IDMA is still an unsolved problem.
The centralized power allocation (CPA) has been tackled by several
authors (see  \cite{Rosberg2007}, \cite{Ping2004}, and references
therein). In these works, iterative multiuser detection/channel
decoding is assumed at the receiver and the base station
determines the power to be transmitted by each user, according to
a global optimum criterion and typically having complete channel
state information (CSI). In this contribution, a novel framework
to tackle the DPA problem in IDMA systems using tools from game
theory is presented.  The proposed approach assumes iterative
multiuser detection/channel decoding at the receiver based on the
chip-by-chip (CBC) algorithm \cite{Ping2006TransWireless}, and
relies on the  signal to noise ratio (SINR) evolution technique
\cite{Ping2004} for non-heavily loaded systems.

In our approach, we assume that each user allocates the power by maximizing its own utility function and by assuming a competitive selfish and rational behavior of the other users. The proposed utility function for a given user is the ratio between a power of the user's goodput (probability of an error-free detected frame) and its own transmit power. Our decentralized approach requires only the knowledge of the noise power at the receiver, the channel gain of the user of interest, and the number of active users in the  system. It yields a channel inversion policy for the power allocation. This policy is applied to an IDMA system with a repetition code of length $N$ bits. Simulations show that the allocated power is substantially independent of the number of users in typical operation conditions for a practical system. Therefore, the knowledge of the number of active users in the system becomes irrelevant as observed in the centralized case in \cite{Rosberg2007}.

\section{System Model}
Consider the uplink of an IDMA system with $K$ chip-synchronous
active users. Each mobile and the base station are equipped with a
single antenna. The base station performs CBC iterative multiuser
decoding based on successive interference cancellation (SIC) or
parallel interference cancellation (PIC) as proposed in
\cite{Ping2006TransWireless}. Denote by $b_k$ the information bits
of the length-$M'$ message to be transmitted by the $k^{th}$ user.
An identical code with low rate $R$ is applied to the messages of
all the users. The coded information bits, referred to as chips,
are permuted by an interleaver $\pi_k$ of length $M=\frac{M'}{R}$
chips. Each interleaver is unique in the network, i.e. $\pi_k \neq
\pi_i$, $\forall i \neq k$. The base band signal $r\left(j\right)$
sampled at the chip-rate at the receiver is
\begin{equation}\label{r}
r\left(j\right) = \displaystyle\sum_{i=1}^{K} \sqrt{p_i}h_ix_i\left(j\right)+n\left(j\right), \; j=0\ldots M-1
\end{equation}
where $j$ is the discrete time index for the chip interval and
$h_i$ and $p_i$ represent the channel realization and the transmit
power of the $i^{th}$ user, respectively. Here,
$x_i\left(j\right)$ represents the chip transmitted by the
$i^{th}$ user at chip interval $j$ and $n\left(j\right)$ is the
additive white Gaussian noise (AWGN) process with zero mean and
variance $\sigma^2$. For the sake of simplicity,  we introduce our
results assuming an antipodal modulation, i.e $x_i\left(j\right)
\in \{-1,1\}$,  real valued channel realizations, and real noise
samples. The extension to more complex modulation schemes is
straightforward. The receiver is made of an elementary multiuser
signal estimator (ESE) based on the principle of maximum a
posteriori (MAP) detection, a set of interleavers $\pi_k $ and
de-interleavers $\pi_k^{-1}$, and a set of $K$ single user a
posteriori probability decoders (DEC). The decoders feed back soft
information to the ESE module in turbo configuration to iterate
and improve the estimations. This structure is known as
CBC decoder and is described in
\cite{Ping2006TransWireless}. The SINR at the input of the decoder can be estimated at each
iteration of the CBC detection by means of the SINR evolution
technique \cite{Ping2004}. It is shown in \cite{Rosberg2007} that
in the steady state, the SINR of user $k$, denoted as $\gamma_k$,
converges to the solution to the following system of equations
\begin{equation}\label{EqSINRSteadyState}
\gamma_{k}^{ss} = \frac{p_k |h_k|^2}{\displaystyle\sum_{i=1,i\neq k}^{K}p_i
|h_i|^2 f\left(\gamma_i^{ss}\right) + \sigma^2}, \quad \forall k \in \left[1,K \right],
\end{equation}
where the super-index $ss$ stands for steady state. The function
$f(x) \in \left[ 0,1\right]$, $\forall x \geqslant 0$ represents
the amount of multiple access interference (MAI) that is
eliminated at each iteration of the CBC detection \cite{Ping2004}.
This function depends on the coding scheme and can be obtained by
Monte Carlo simulations as described in
\cite{Ping2006TransWireless}. The system (\ref{EqSINRSteadyState})
might have several solutions. Nonetheless, it has been shown in
\cite{Rosberg2007} that in the case of $|h_i|^2p_i = |h_j|^2p_j$
$\forall i \neq j$, the system has a unique solution given by the
fix point equation
\begin{equation}\label{EqSINRSSUniqueSolution}
\gamma^{ss} = \frac{p_k |h_k|^2}{(K-1)|h_k|^2p_k
f\left(\gamma^{ss}\right) + \sigma^2} \quad \forall k \in \left[1,K \right]
\end{equation}
and $\gamma_k^{ss} = \gamma^{ss}$, with $1 \leqslant k \leqslant K$.

Additionally, it has been shown that the SINR evolution technique
predicts precisely the SINR under the constraint of non-overloaded
systems \cite{Ping2006TransWireless}, \cite{Ping2004}. However, we
have found in our simulations that this is not the case for
heavily overloaded systems. Under this constraint, the SINR
evolution technique does not match the SINR obtained at the output
of the ESE module. In order to exclude this case, we restrict our
study to non-overloaded systems, i.e. we consider that
$\frac{K}{N} \leqslant 1$ always holds.
\section{A Non-Cooperative Power Allocation Game}\label{sec_game}
We define the power allocation problem as a strategic game denoted
by the triplet $\left\lbrace
\mathrm{S},\mathcal{P},\left(u_k\right)_{k \in
\mathrm{S}}\right\rbrace$, where $\mathrm{S} = \left\lbrace
1,\ldots, K\right\rbrace$ is the set of players or active mobiles,
$\mathcal{P}$ represents the set of strategies, and $u_k$ is the
utility function of user $k$. In this work, the strategy of each
user consists of all the possible transmittable power levels $ p_k
\in \mathbb{R}^+$, $\forall k \in \mathrm{S}$. We follow the same
line as in \cite{Rosberg2007}, \cite{Bonneau2007} and
\cite{Meshkati2005} and enforce constraints on the maximum
transmit power later. Furthermore, the utility function $u_k$ of
user $k$ is defined as the ratio between the positive power $s_k$
of its goodput $g(\gamma_k),$ i.e. the probability of an
error-free detected packet, and its transmit
power\footnote{Further discussions on the definition of this
utility function in $p_k=0$ can be found in \cite{Bonneau2007} and
\cite{Meshkati2005}.}
\begin{equation}\label{EqUtilityFunction}
u_k\left(\mathbf{p},\mathbf{h}\right) =
\frac{\left( g\left(\gamma_k\right)\right)^{s_k}}{p_k} \quad \forall k \in \mathrm{S},
\end{equation}
where $\boldsymbol{p}=\left(p_1, p_2,\ldots,p_K\right)$ and
$\boldsymbol{h}=\left(h_1, h_2,\ldots,h_K\right)$ are the vectors
of transmit power levels and channel realizations, respectively.
Expression (\ref{EqUtilityFunction}) generalizes the utility
function presented in \cite{Bonneau2007} and \cite{Meshkati2005}
when $s_k=1$, $\forall k \in \mathrm{S}$. With respect to the
reference utility function with $s_k=1$ the user $k$ will choose
$s_k>1$ if it is interested in better performance at the expenses
of a higher transmitted power. On the contrary, if it needs to
save power then it will set $s_k<1.$ In this study, we assume that
all the users are aware of the fact that they are all interested
on the same  QoS profile, i.e. $s_1=\ldots=s_K = s$.

The objective of the player $k$ is to determine the  transmit power $p_k^*$ that selfishly maximizes its utility function $u_k(\mathbf{p}^*,\mathbf{h})$ under the assumption that a similar strategy is adopted by all the other users. Thus, the optimum power \cite{Perlaza2007MasterThesis} is obtained as solution to the following system of equations:
\begin{equation}\label{EqDerivativeUtility}
 \begin{array}{cl}
 \frac{\partial}{\partial p_k}u_k(\boldsymbol{p}^*,\boldsymbol{h})&= \left(\frac{g(\gamma_k)^{s-1}}{(p_k)^2}\right)\left(s g'(\gamma_k) p_k  \frac{\partial}{\partial p_k} \gamma_k - g\left(\gamma_k\right)\right) \\  &= 0\quad \forall k \in \mathrm{S}.
 \end{array}
\end{equation}
Here, $g(\gamma_k)'$ represents the first derivative of $g(\gamma_k)$.  Due to the nature of the channel (non-error free channel), the term $\frac{g(\gamma_k)^{s-1}}{(p_k^*)^2}$, $\forall k \in \mathrm{S}$ is always non-zero. Thus, the system in (\ref{EqDerivativeUtility}) could be simplified as
\begin{equation}\label{EqUtilityFunction_Derivative}
 s g'(\gamma_k) p_k  \frac{\partial}{\partial p_k} \gamma_k - g\left(\gamma_k\right) = 0 \quad \forall k \in \mathrm{S}. \end{equation}
Furthermore, it has been shown in \cite{Perlaza2007MasterThesis} that
\begin{equation}\label{SINR_Derivative}
\frac{\partial}{\partial p_k}\gamma_k = \frac{\gamma_k}{p_k} \left( \frac{1+\gamma_k f\left(\gamma_k\right) } {1-\gamma_k^2 f'\left(\gamma_k \right)}\right) q_k(\boldsymbol{\gamma}),\; \forall k \in \mathrm{S}
\end{equation}
where $\boldsymbol{\gamma} = \left(\gamma_1,\ldots,\gamma_K\right)$ and
\begin{equation*}
q_k(\boldsymbol{\gamma}) = \left(1 - \frac{\gamma_kf\left(\gamma_k\right) +  \gamma_k^2 f'\left(\gamma_k\right) }  {\left(1-\gamma_k^2f'\left(\gamma_k\right)\right)\left(1+\gamma_kf\left(\gamma_k\right)\right)\left( A + 1\right) }\right)
\end{equation*}
and
\begin{equation*}
A = \displaystyle\sum_{i=1}^{K} \frac{\gamma_i^2 f'\left(\gamma_i\right)}{\left(1-\gamma_i^2 f'\left(\gamma_i\right)\right)}.
\end{equation*}
Replacing (\ref{SINR_Derivative}) in (\ref{EqUtilityFunction_Derivative}) we obtain the system of equations
\begin{equation}\label{EqSystemOfEquations}
s g'\left(\gamma_k\right) \gamma_k \left( \frac{1+\gamma_k f\left(\gamma_k\right) } {1-\gamma_k^2 f'\left(\gamma_k \right)}\right) q_k(\boldsymbol{\gamma}) - g\left(\gamma_k\right) = 0, \forall k  \in \mathrm{S}
\end{equation}
In (\ref{EqSystemOfEquations}), the variables are only $\gamma_k$,
$\forall k \in \mathrm{S}$. Therefore, the system does not depend
directly on the channel realizations and the transmit power of
each user. This property was also observed in the centralized
approach \cite{Rosberg2007}. Moreover, in the system
(\ref{EqSystemOfEquations}), all the equations are identical and
the system is invariant to variable permutations. This implies
that the solutions are also identical, i.e. $\gamma_1 =
\gamma_2=\cdots= \gamma_K = \gamma$. Thus, the system
(\ref{EqSystemOfEquations}) reduces to a single equation
\begin{equation}\label{EqTargetFunction}
s g'\left(\gamma \right) \gamma  q(\gamma) - g\left(\gamma \right) = 0
\end{equation}
where
\begin{equation*}
\begin{array}{cl}
q(\gamma) = & \frac{\left( 1-\gamma^2 f'\left( \gamma \right) \right) + \left( K-1 \right) \left( \gamma^2 f'\left( \gamma \right) \right) \left( 1+\gamma f\left( \gamma \right) \right)}{\left( 1-\gamma^2 f'\left( \gamma \right) \right) \left(1 + \left( K-1 \right)  \gamma^2 f'\left( \gamma \right) \right)}.
\end{array}
\end{equation*}

We define $z(\gamma)=s g'\left(\gamma \right) \gamma  q(\gamma) - g\left(\gamma \right)$ and we name it \emph{target function}. The zeros of the target function are candidates to be the optimal SINR $\gamma^*$. The optimal SINR corresponds to the SINR at which the utility function of each user is maximized. If the set of optimal SINR $\gamma_i^* = \gamma^*$, $\forall i \in \mathcal{S}$ is known, it is possible to obtain an expression for the optimal transmit power level from expressions (24) and (25) in \cite{Rosberg2007},
\begin{equation}\label{EqOptimalPower}
p_k^* = \frac{1}{|h_k|^2} \left(\frac{\sigma^2
\gamma^*}{1-\left(K-1\right)\gamma^*
f\left(\gamma^*\right)}\right) \quad \forall k.
\end{equation}
Eqn. (\ref{EqOptimalPower}) holds under the
constraint\footnote{The constraint is widely satisfied since the
function $f(\gamma)$ decreases rapidly with $\gamma$
\cite{Ping2006TransWireless}, \cite{Rosberg2007}}
\begin{equation}\label{Kmax}
 K < \left \lfloor \frac{1}{\gamma^* f\left(\gamma^*\right)} + 1 \right \rfloor
\end{equation}
since only positive power levels are meaningful. Note that the expression (\ref{EqOptimalPower}) could be also obtained from the fix point equation (\ref{EqSINRSSUniqueSolution}) as unique solution.

Interestingly, the power allocation for a given user depends on the optimal SINR $\gamma^*,$ its channel gain, and the number of active users.  Therefore, the knowledge of all the other users' channel gains in the network is not required.

\section{A Decentralized Power Allocation Algorithm}\label{SecAlgorithm}
In this section, we propose a power allocation algorithm based on
the game investigated in Section \ref{sec_game} under the
constraint of non-overloaded systems, i.e.
$\frac{K}{N}\leqslant1$. 
In general, the target function $z(\gamma)$ is not linear and the
solution to $z(\gamma)=0$ requires a numerical approach. The
algorithm we present here is based on the secant method. The
iterative search of the optimum SINR $\gamma^*$ is initialized by
choosing two values $\gamma_1$ and $\gamma_2$ such that
\begin{itemize}
  \item both $\gamma_1$ and $\gamma_2$ are not lower than the
  minimum SINR required $\gamma_{\mathrm{min}}$ for reliable communications at rate
  $R$ without considering SIC or PIC. According to the Shannon capacity law  $C=
  \frac{1}{2}\log_2(1+\gamma)$. Thus, $\gamma_{\mathrm{min}}=2^{2R}-1$ and $\gamma_{\mathrm{min}}\leq\gamma_1 \leq \gamma_2;$
  \item $z(\gamma_2)\leq z(\gamma_{1}).$
\end{itemize}

Furthermore, in practical systems, the users have power
constraints. It might happen that the optimal power $p_k^*$ for a
given optimal SINR $\gamma^*$ exceeds the maximum transmittable
power $p_{\mathrm{max}}$. In this case, a user $i$, $\forall i \in
\mathcal{S}$ could either transmit at the maximum power $p_i =
p_{\mathrm{max}}$ or not transmit $p_i = 0$ \cite{Rosberg2007}.
Let us denote as
$\gamma_{k,0}^{\mathrm{p_{max}}}=\frac{p_{max}|h_k|^2}{(K-1)p_k|h_k|^2
+ \sigma^2}$ the SINR achieved by the $i^{th}$ user before the
first iteration of the CBC algorithm when transmitting at the
maximum power $p_{\mathrm{max}}.$  Indeed, if
$\gamma_{k,0}^{\mathrm{p_{max}}}$   enables reliable decoding,
i.e. $\gamma_{k,0}^{p_{\mathrm{max}}} \geqslant
\gamma_{\mathrm{min}}$, then the user transmits at the maximum
power level $p_{\mathrm{max}}$. In this case, the user does not
reach the optimal $\gamma^*$. However, a reliable decoding is always ensured.

If the condition $\gamma_{k,0}^{p_{\mathrm{max}}} \geqslant
\gamma_{\mathrm{min}}$ is not satisfied, the $k^{th}$ user does
not transmit and is considered in \emph{outage}. Note that, the
condition over the SINR $\gamma_{k,0}^{p_{\mathrm{max}}}$ is a
necessary but not sufficient condition for a user  not to be
decoded. In fact, certain users could attain an SINR
$\gamma_{k,0}^{p_{\mathrm{max}}}$  higher than
$\gamma_{\mathrm{min}}$ after decoding as a result of the
iterative decoding. However, a user cannot evaluate this
possibility. In fact, due to the incomplete knowledge available
about other users, it can not determine if iterative decoding is
able to sufficiently improve the initial SINR and enable a
reliable decoding. Then, transmitting might result in a waste of
energy and additional interference for all the other users.

Therefore, the power allocation rule is
\begin{equation}\label{EqPowerAllocationRule}
 p_k = \left\{  \begin{array}{lcl}
 p_k^*& \text{if} &  p_{k}^* \leq p_{\mathrm{max}}\\
 p_{max} & \text{if} & p_k^{*} > p_{max}  \text{ and }  \gamma_{k,0}^{\mathrm{p_{max}}} \geq \gamma_{min}\\
 0 & \text{otherwise} &
 \end{array}\right.
 \end{equation}
Let us denote with $\varepsilon > 0$ the desired accuracy to determine $\gamma^*.$ The power allocation algorithm is summarized as follows
\begin{enumerate}
\item \emph{Initialization of the secant method}\\
$\gamma_1 = \gamma_{min}$ \\
\texttt{while} ($z\left( \gamma_1 \right)  \geqslant z\left( \gamma_{min} \right)$) \texttt{then}\\
    $\gamma_1 = \gamma_{1} + \Delta \quad \text{with } \Delta \in \left(0,\frac{1}{2}\right]$\\
\texttt{end}\\
$\gamma_2 = \gamma_{1} + \Delta \quad \text{with } \Delta \in \left(0,\frac{1}{2}\right]$
\item \emph{Iterative step of the secant method}\\
\texttt{while} $|\gamma_{i+1}-\gamma_{i}|>\varepsilon$ \\ $
\gamma_{i+1} = \gamma_{i} - \frac{\gamma_i -
\gamma_{i-1}}{z\left(\gamma_i\right) - z\left(\gamma_{i-1}\right)}
z\left( \gamma_i\right)$\\ \texttt{end}
\item \emph{End of the secant method}\\
$\gamma^* = \gamma_{i+1}$.
\item \emph{Power allocation}
Determine the transmit power according to
(\ref{EqPowerAllocationRule}).
\end{enumerate}
\section{Performance Assessment}
The performance  of the proposed power allocation scheme is assessed assuming antipodal modulation and a repetition code with different rates $R=\frac{1}{N}$ and $N=$ $16,$ $32,$ $64$. The channel gains and the noise are real and Gaussian distributed with zero mean and unit variance.

We refer to the SINR at which the utility function
(\ref{EqUtilityFunction}) is maximized as the optimal operating
point $\gamma^*$. An approximate expression for the bit error rate
when a repetition code with rate $R=\frac{1}{N}$ is used is
\begin{equation}\label{Pe_approx}
\begin{array}{cl}
P_e \thickapprox & \displaystyle\sum_{i = 0}^{\left\lfloor \frac{N}{2}\right\rfloor} \left(\begin{array}{c} N \\ \left\lceil \frac{N}{2} \right\rceil +i\end{array} \right) \left( Q\left(\sqrt{2\gamma_k}\right) \right)^{\left\lceil \frac{N}{2} \right\rceil+i} \\
&  \left( 1 - Q\left(\sqrt{2\gamma_k}\right) \right)^{\left\lceil \frac{N}{2}\right\rceil -i }
\end{array}
\end{equation}
In (\ref{Pe_approx}), the function $Q(.)$ is the complementary
error function of a Gaussian random variable. It represents the
probability of error of one chip in the case of antipodal
modulation. Hence, the goodput can be written as:
\begin{equation}\label{Functiong_Final}
g\left(\gamma_k\right) = \left(1-P_e\right)^M.
\end{equation}
For the coding scheme considered in this contribution, the
function $f(\cdot)$ is determined by Monte Carlo simulations as
described in \cite{Ping2006TransWireless}. The obtained results
 match those presented in \cite{Rosberg2007} and
\cite{Lau2007} perfectly.

\begin{figure}
\begin{center}
\includegraphics[width=\linewidth]{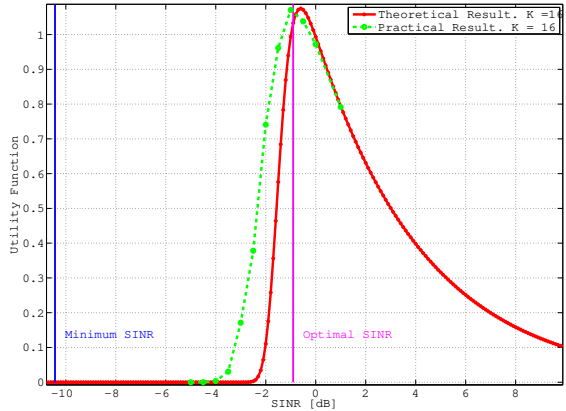}
\caption{\label{FigUtilityFunction} Average Utility Function versus the SINR in decibels for a standard IDMA system with $K=16$ users, $M = 1000$ bits and coding rate $R=\frac{1}{16}$. Solid and dashed lines correspond to the SINR evolution and the system simulation, respectively.}
\end{center}
\end{figure}

\begin{figure}
\begin{center}
\includegraphics[width=\linewidth]{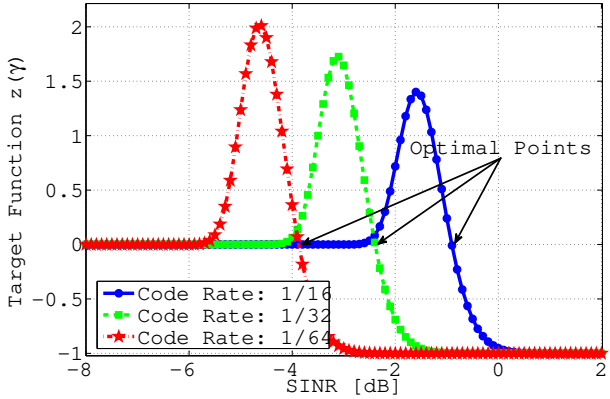}
\caption{\label{FigTargetFunction_SeveralRates} Target function $z(\gamma)$ versus SINR in decibels for a standard IDMA system with parameters $K = 16$ users, $M = 1000$ bits. The solid, dot-dashed and dashed lines correspond to the coding rates  $R=\frac{1}{16}$, $R=\frac{1}{32}$ and $R=\frac{1}{64}$ respectively.}
\end{center}
\end{figure}

\begin{figure}
\begin{center}
\includegraphics[width=\linewidth]{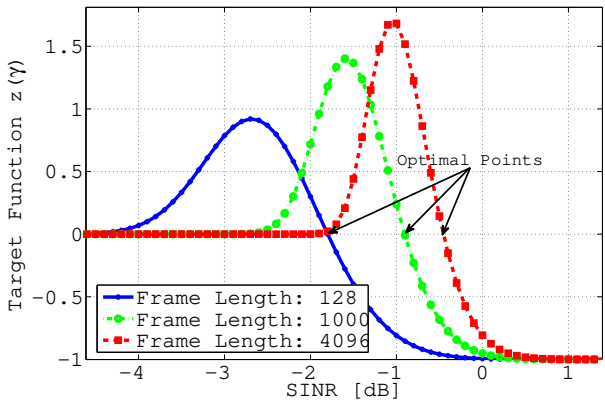}
\caption{\label{FigTargetFunction_SeveralFrameLengths} Target function $z(\gamma)$ versus SINR in decibels for a standard IDMA system with parameters $K = 16$ users, coding rate $R = \frac{1}{16}$ bits. The solid, dot-dashed and dashed  lines correspond to the frame lengths $128$, $1000$ and $4096$ bits respectively.}
\end{center}
\end{figure}

\begin{figure}
\begin{center}
\includegraphics[width=\linewidth]{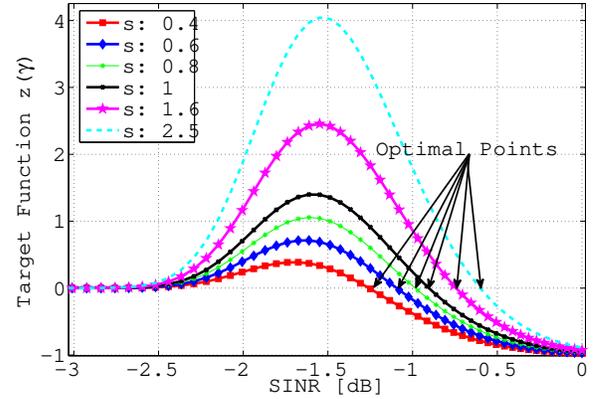}
\caption{\label{FigTargetFunction_Severals} Target function $z(\gamma)$ versus SINR in decibels for a standard IDMA system with coding rate $R = \frac{1}{16}$ , frame length $M = 1000$ bits, $K = 16$ users and different values for the parameter $s$}
\end{center}
\end{figure}

\begin{figure}
\begin{center}
\includegraphics[width=\linewidth]{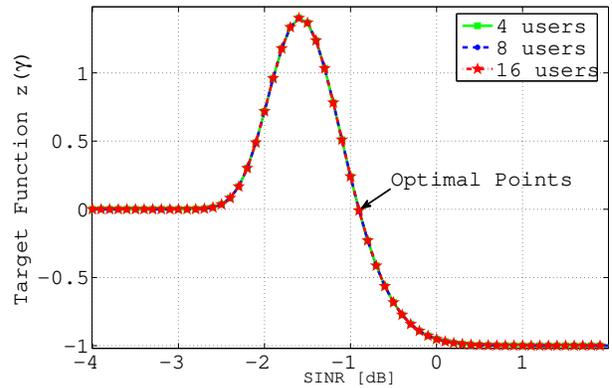}
\caption{\label{FigTargetFunction_SeveralUsers} Target function $z(\gamma)$ versus SINR in decibels for a standard IDMA system with coding rate $R = \frac{1}{16}$ , frame length $M = 1000$ bits. The solid, dashed, and dot-dashed lines correspond to $K = 4 $, $K = 8$ and $K = 16$ users respectively.}
\end{center}
\end{figure}

In Figure \ref{FigUtilityFunction}, we plot the average utility
function versus the SINR $\gamma$ when $s=1$. The solid line shows
the theoretical performance determined by plugging
(\ref{Functiong_Final}) in (\ref{EqUtilityFunction}), while the
dashed lines are obtained with the actual BER at the output of the
decoder. We  noticed in all the non-overloaded cases that the
utility has a unique maximum.
The effects of the coding rate, the frame length, and the number
of users on the optimum operating point are assessed in Figures
\ref{FigTargetFunction_SeveralRates},
\ref{FigTargetFunction_SeveralFrameLengths},
\ref{FigTargetFunction_Severals}, and
\ref{FigTargetFunction_SeveralUsers}. In Figure
\ref{FigTargetFunction_SeveralRates}, the target function
(\ref{EqTargetFunction}) is plotted as a function of the SINR
$\gamma$ for three different coding rates while the frame length
and the number of users are kept constant. Lower coding rates
determine lower optimal SINRs. In Figure
\ref{FigTargetFunction_SeveralFrameLengths}, the target function
is plotted as a function of the SINR $\gamma$ considering three
different frame lengths,  keeping the coding rate and the number
of users constant. In this case, longer frames lead to higher
optimal SINR values. In Figure \ref{FigTargetFunction_Severals}, we
plot the target function as a function of the SINR for several
exponents  $s$ of the generalized utility function
(\ref{EqUtilityFunction}), while keeping the number of users, the
frame length and the coding rate constant. When users are
interested in better performance rather than saving power $(s>1)$,
the optimal SINR increases and vice versa. In Figure
\ref{FigTargetFunction_SeveralUsers}, the target function is
plotted for 4, 8,  and 16 active users as a function of the SINR
when the coding rate and frame length are kept constant. The four
lines overlap completely. Therefore, the variation of the number
of active users $K$ has a negligible effect on the optimal SINR.
Then, the optimal operating point is practically independent of
the number of users. This effect was also observed in the
centralized case in \cite{Rosberg2007}.

In general, a convenient solution for a non-cooperative game is
the Nash equilibrium (NE). By definition, the NE is a solution
such that no player is interested on changing its strategy since
no improvement could be obtained in its own utility while keeping
the other users' utilities unchanged. In this case, even though
the simulations results show that there is a unique solution to
the game $(\gamma^*)$, we could not prove analytically the
uniqueness of the NE since the functions $f(\gamma)$ and
$g(\gamma)$ depend on the coding scheme. However, particular cases
have been already studied. In the case where no interference
cancellation is performed, i.e $f(\gamma) = 1$, the proof is
similar to the one presented in \cite{Bonneau2007} with $s=1$ for
the CDMA case.


\section{Conclusions}
We provided a novel framework for decentralized power allocation in IDMA systems based on a game-theoretic approach. In this context, each user aims at  selfishly maximizing its own utility function assuming a similar behavior is adopted by all the other users. It leads to a channel inversion power allocation policy where the optimal power level could be set at each terminal based on the knowledge of its own channel realization, the noise energy at the receiver, and the number of active users in the network. Interestingly, we found that under practical operation conditions, the knowledge of the number of active users becomes irrelevant as observed in the centralized case in \cite{Rosberg2007}.
\section{Acknowledgements}
This work is developed in the frame of \emph{Tamara project} supported by \emph{Groupe des Ecoles des T\'el\'ecommunications (GET, France)}. Samir Medina Perlaza is supported by \emph{Programme Al$\beta$an}, the European Union programme of high level scholarships for Latin America, scholarship No. E06M101130CO.
\bibliographystyle{IEEEtran}
\bibliography{samedina}
\end{document}